\documentstyle[preprint,floats,prd,aps]{revtex}
\tightenlines 
\begin{document}
%
\preprint{\vbox {\hbox{OCHA-PP-134}}}

\draft
\title{A more careful estimate of the charm content of $\eta'$} \author{Mohammad R. Ahmady$^a$
\footnote{Email: mahmady2@julian.uwo.ca}
, Emi Kou$^b$\footnote{Email: g9870407@edu.cc.ocha.ac.jp} }
\address{
$^a$Department of Applied Mathematics\\ University of Western Ontario\\ London, Ontario N6A 5B7, Canada \\
$^b$ Department of Physics, Ochanomizu University \\
1-1 Otsuka 2, Bunkyo-ku,Tokyo 112-0012, Japan}

\date{March 2000}
\maketitle
\begin{abstract}
We estimate the quantity $\vert f_{\eta'}^{(c)}\vert$ which is associated with the charm content of $\eta'$ meson from the experimentally known ratio $R=B (\psi\to\eta'\gamma )/B (\psi\to\eta_c\gamma )$.  It is shown that due to the off-shellness of the $c\bar c$ component of $\eta'$, which has been overlooked so far, $f_{\eta'}^{(c)}$ is further suppressed.  Assuming that $\psi\to\eta'\gamma$ decay is dominated by $\psi\to\eta_c$ transition, we obtain $\vert f_{\eta'}^{(c)}\vert \approx 2.4$ MeV which could imply that the $b\to c\bar c s$ mechanism does not play a major role in the $B\to K\eta'$ decay mode.
\end{abstract}
%

\newpage
Various properties of $\eta'$ meson have been at the focus of a lot of theoretical attentions.  Recently, a fresh interest in this psuedoscalar particle has arisen due to the measurement of unexpectedly large branching ratios for inclusive $B\to X_s\eta'$ and exclusive $B\to K\eta'$ decay modes by the CLEO collaboration\cite{s,b,bw}.  There have been various attempts at explaining these experimental results within or beyond the Standard Model.  For example, anomalous coupling of $\eta'$ to two gluons has been used in conjunction with the QCD penguin to reproduce the observed results\cite{as,aks}.  On the other hand, it has been argued that the possible charm content of $\eta'$ plus the the CKM favored $b\to c\bar cs$ transition could be responsible for the large $\eta'$ production in B meson decays\cite{hz}.

In this work, we investigate whether or not $\eta'$ contains a sizable charm component.  The parameter $f_{\eta'}^{(c)}$ which is defined as
\begin{equation}
\langle 0\vert\bar c\gamma_\mu\gamma_5 c\vert\eta'(q)\rangle =f_{\eta'}^{(c)}q_\mu\;\; ,
\end{equation}
is estimated by utilizing the observed value for the ratio $R=B (\psi\to\eta'\gamma )/B (\psi\to\eta_c\gamma )$.  For this purpose, one can write the $\eta'$ meson state in terms of its various possible components
\begin{equation}
\vert\eta'\rangle =C_1\vert\eta_1\rangle +C_8\vert\eta_8\rangle +C_g\vert gg\rangle +C_c\vert\eta_c\rangle +...\;\; ,
\end{equation}
where $\vert\eta_1\rangle$ and $\vert\eta_8\rangle$ are flavor $SU(3)$ singlet and octet states, respectively, and $\vert gg\rangle$ represents a glueball state.  The last term in Eq. (2) is the $c\bar c$ content of $\eta'$ which should have the same quantum numbers as $\eta_c$.  The probability amplitude of finding  $\vert\eta'\rangle$ in any of its components is described by the coefficients $C_i$, $i=1,8,g,c$ in Eq. (2).  Here an explanation about the inclusion of the gluon and charm components that may appear due to the $U(1)_A$ anomaly, is in order.  The role of the strong anomaly in the low energy dynamics of the $\eta'$ meson was established by $^,$t Hooft\cite{th}, Witten\cite{witten} and Veneziano\cite{veneziano}.  In fact, one can write a low energy effective chiral Lagrangian for the meson field which obeys the anomalous conservation law\cite{rst,vv,na} and where other degrees of freedoms (like glueballs etc.) are integrated out (or equivalently, eliminated by using the equations of motion).  Therefore, this effective Lagrangian may be expressed purely in terms of the light meson fields\cite{miransky} which is useful if we are interested only in $\eta'$ meson.  However, to examine various mechanisms in the fast $\eta'$ production in two body B decays, the conventional approach is to write all possible states that mix with this anomalous psuedoscalar meson explicitly.  The mixing coefficients, i.e. $C_i$, are in principle related if they are calculated from the underlying dynamics.  However, here they are considered as phenomenological parameters to be determined from experimental data.

From Eqs. (1) and (2), to leading order in $1/m_c$, one obtains 
\begin{eqnarray}
\nonumber f_{\eta'}^{(c)}q_\mu &=&C_c\langle 0\vert\bar c\gamma_\mu\gamma_5 c\vert\eta_c(q)\rangle \\
&=& C_cf_{\eta_c}(q^2=m_{\eta'}^2)q_\mu\;\; ,
\end{eqnarray}
which results in
\begin{equation}
f_{\eta'}^{(c)}=C_cf_{\eta_c}(q^2=m_{\eta'}^2)\;\; .
\end{equation}
We note that $q$ is the momentum of the physical $\eta'$ meson and hence, $f_{\eta_c}$ should be evaluated far off $\eta_c$ mass-shell as is explicitly shown in Eqs. (3) and (4).  This important issue has not been taken into account so far in the estimates of $f_{\eta'}^{(c)}$ and is the main point of the present work.  In fact, we show that the off-shellness effect leads to the suppression of $f_{\eta_c}$ and, consequently, a smaller value for $f_{\eta'}^{(c)}$ is obtained. 

The value of on-mass-shell $f_{\eta_c}$ is extracted from the two photon decay rate of $\eta_c$
\begin{equation}
\Gamma (\eta_c\to\gamma\gamma )=\frac{4{(4\pi\alpha )}^2f_{\eta_c}^2(m_{\eta_c}^2)}{81\pi m_{\eta_c}}\;\; .
\end{equation}
Using the measured decay width $\Gamma (\eta_c\to\gamma\gamma )=7.5^{+1.6}_{-1.4}$ KeV\cite{pdg} results in an estimate of $f_{\eta_c}(m_{\eta_c}^2)=411$ MeV where $m_{\eta_c}^2$ in the parentheses is to emphasize that the obtained number is for on-mass-shell $\eta_c$.  However, as it is pointed out in Ref. \cite{aks2}, a model calculation of $\eta_c$-photon-photon coupling reveals a drastic suppression of the $\eta_c\to\gamma\gamma$ transition form factor $g(q^2)$ when $q^2$ is small compared to its on-shell value, i.e. $q^2\ll m_{\eta_c}^2$.  In this model, the two photon decay of $\eta_c$ proceeds via a triangle quark loop which is illustrated in Fig. 1.  The corresponding expression can be written in the following form
\begin{equation}
T^{\mu\nu}(\eta_c\to\gamma\gamma )=Ng(q^2)\epsilon^{\mu\nu\alpha\beta}p_{1\alpha}p_{2\beta} \;\; ,
\end{equation}
where $p_1$ and $p_2$ are the four-momenta of the photons and $q=p_1+p_2$.  The form factor $g(q^2)$ is obtained from the quark loop calculation: 
\begin{equation}
g(q^2)\begin{array}[t]{l}\displaystyle = \int^1_0dx\int_0^{1-x}dy\frac{1}{m_c^2-q^2xy}  \\
=\left  \{ { \begin{array}{l}
\displaystyle\frac{-2}{q^2}Arcsin^2\sqrt{\frac{q^2}{4m_c^2}}\;\;\; 0\le q^2\le 4m_c^2 \\
\displaystyle\frac{2}{q^2}{\left [ Ln\left (\sqrt{\frac{q^2}{4m_c^2}}+\sqrt{\frac{q^2}{4m_c^2}-1}\right )-\frac{I\pi}{2}\right ]}^2 \;\;\; 4m_c^2\le q^2\end{array}}\right . \;\; , 
\end{array}
\end{equation}
where $m_c$ is the charm quark mass.  In Fig. 2, the variation of $g(q^2)/g(m_{\eta_c}^2)$ in the range $m_{\eta'}^2\le q^2\le m_{\eta_c}^2$ is depicted.  We observe that for $q\approx m_{\eta'}^2$, the form factor suppression is quite substantial.  In writing Eq. (6), the constants are all swept into the factor $N$ which can be obtained using the requirement that for $q^2=m_{\eta_c}^2$ Eq. (6) should yield the experimentally measured decay rate $\Gamma (\eta_c\to\gamma\gamma )$.  Consequently, we obtain the following form for the $\eta_c$-$\gamma\gamma$ transition amplitude: 
\begin{equation} \displaystyle 
A(\eta_c\to\gamma\gamma)= \frac{16i\sqrt{m_{\eta_c}\Gamma (\eta_c\to\gamma\gamma )}}{\pi^{3/2}}g(q^2)\epsilon^{\mu\nu\alpha\beta}\epsilon_\mu (p_1)\epsilon_\nu (p_2)p_{1\alpha}p_{2\beta}\;\; . \end{equation} $\epsilon (p_i)$ is the polarization of the photon with momentum $p_i$ and we assumed weak binding for charmonium, i.e. $m_{\eta_c}\approx 2m_c$.  Eqs. (5) and (8) lead to the following result
\begin{eqnarray}
 \nonumber f_{\eta_c}(q^2=m_{\eta'}^2)&=&\displaystyle\frac{g(m_{\eta'}^2)}{g(m_{\eta_c}^2)}f_{\eta_c}(m_{\eta_c}^2) \\
&=&\displaystyle\frac{m_{\eta_c}^2}{m_{\eta'}^2}\frac{{Arcsin}^2\sqrt{\frac{m_{\eta'}^2}{m_{\eta_c}^2}}}{{(\frac{\pi}{2})}^2}f_{\eta_c}(m_{\eta_c}^2)\;\; ,
\end{eqnarray}
where the last term is obtained by using Eq. (7).  As a result, we observe that $f_{\eta_c}$ on $\eta'$ mass-shell
\begin{equation}
f_{\eta_c}(q^2=m_{\eta'}^2)\approx 0.42f_{\eta_c}(m_{\eta_c}^2)\approx 172 \;{\rm MeV}\;\; ,
\end{equation}
is reduced to less than 50\% of its value for on-mass-shell $\eta_c$. 

To proceed with the numerical estimate of $f_{\eta'}^{(c)}$ via Eq. (4), we use the branching ratios $B(\psi\to\eta'\gamma )=(4.31\pm 0.30)\times 10^{-3}$ and $B(\psi\to\eta_c\gamma )=(1.3\pm 0.4)\times 10^{-2}$ which are experimentally known\cite{pdg}.  Assuming that the former decay mode dominantly occurs through $\psi$ transition to the $\eta_c$ component of $\eta'$ results in
\begin{equation}
R=\frac{B(\psi\to\eta'\gamma )}{B(\psi\to\eta_c\gamma )}=C_c^2\frac{{(m_\psi^2-m_{\eta'}^2)}^3}{{(m_\psi^2-m_{\eta_c}^2)}^3}\;\; .
\end{equation}
We evaluate $C_c$ by inserting the central value of the branching ratios in Eq. (11) which yields
\begin{equation}
\vert C_c\vert =0.014\;\; ,
\end{equation}
and consequently, leads to our estimate for $\vert f_{\eta'}^{(c)}\vert$
\begin{equation}
\vert f_{\eta'}^{(c)}\vert \approx 2.4\; {\rm MeV}\;\; .
\end{equation}
We note that the stringent bound in Eq. (12) is considerably lower than the estimated range of (50-180) MeV for $f_{\eta'}^{(c)}$ in Refs. \cite{hz} and \cite{sz}.\footnote{Some recent estimates along the same line point to smaller results\cite{fppg,amt}}  The value of $\vert f_{\eta'}^{(c)}\vert$ obtained by us is less than half of the estimates in Refs. \cite{ag} and \cite{fk} due to the fact that the off-shellness effect of the $c\bar c$ component of $\eta'$ has been taken into account in our evaluations.  At the same time, the estimate given in Eq. (12) is within the range $-65\; {\rm MeV}\le f_{\eta'}^{(c)}\le 15\; {\rm MeV}$ presented in Ref. \cite{fk2} based on an analysis of the transition form factor data which is also consistent with $f_{\eta'}^{(c)}=0$.

In conclusion, we estimated the parameter $f_{\eta'}^{(c)}$, which is related to the charm content of $\eta'$, by using experimental inputs and considering the fact the pseudoscalar $c\bar c$ component of $\eta'$ is highly off mass-shell.  Our stringent bound could imply that the decay mode $B\to K\eta'$ does not receive significant contribution from $b\to c\bar cs$ transition.

\vskip 2cm
\noindent
{\bf\Large Acknowledgement}\\
 We would like to thank V. A. Miransky and V. Elias for useful discussions.  M. A. acknowledges support from the Science and Technology Agency of Japan.  E. K. acknowledges support from the Japanese Society for the Promotion of Science.


\begin{references}

\bibitem{s} J. G. Smith, preprint COLO-HEP-395 (1998), hep-ex/9803028.

\bibitem{b} B. H. Behrens et al. (CLEO Collaboration), Phys. Rev. Lett. {\bf 80}, 3710 (1998).

\bibitem{bw} T. E. Browder et al. (CLEO Collaboration), preprint CLNS 98/1544, CLEO 98-4, hep-ex/9804018. 

\bibitem{as} D. Atwood and A. Soni, Phys. Lett. B {\bf 405}, 150 (1997).

\bibitem{aks} M. R. Ahmady, E. Kou and A. Sugamoto, Phys. Rev. D {\bf 58}, 014015 (1998).

\bibitem{hz} I. Halperin and A. Zhitnitsky, Phys. Rev. Lett. {\bf 80}, 438 (1998); Phys. Rev. D {\bf 56}, 7247 (1997). 

\bibitem{th}  G. $^,$t Hooft, Phys. Rev. Lett. {\bf 37}, 8 (1976).

\bibitem{witten} E. Witten, Nucl. Phys. B {\bf 156}, 269 (1979).

\bibitem{veneziano} G. Veneziano, Nucl. Phys. B {\bf 159}, 213 (1979).

\bibitem{rst} C. Rosenzweig, J. Schechter and C. G. Trahern, Phys. Rev. D {\bf 21} 3388 (1980).

\bibitem{vv} P. Di Vecchia and G. Veneziano, Nucl. Phys. B {\bf 171}, 253 (1980).

\bibitem{na} P. Nath and R. Arnowitt, Nucl. Phys. B {\bf 209}, 234 (1982). 

\bibitem{miransky} For a brief insight see V. A. Miransky, {\it Dynamical symmetry breaking in quantum field theories} (World Scientific, 1993) section 12.12.

\bibitem{pdg} Particle Data Group, C. Caso et al., Eur. Phys. J. C {\bf 3}, 1 (1998).

\bibitem{aks2} M. R. Ahmady, E. Kou and A. Sugamoto, Phys. Rev. D {\bf 57}, 1997 (1998).

\bibitem{sz} E. V. Shuryak and A. R. Zhitnitsky, Phys. Rev. D {\bf 57}, 2001 (1998).

\bibitem{ag} A. Ali and C. Greub, Phys. Rev. D {\bf 57}, 2996 (1998).

\bibitem{fk} T. Feldmann, P. Kroll and B. Stech, Phys. Rev. D {\bf 58}, 114006 (1998).    

\bibitem{fk2} T. Feldmann and P. Kroll, Eur. Phys. J. C {\bf 5}, 327 (1998).


\bibitem{fppg} M. Franz, P. V. Pobylitsa, M. V. Polyakov and K. Goeke, hep-ph/9810343.

\bibitem{amt} F. Araki, M. Musakhanov and H. Toki, Phys. Rev. D {\bf 59}, 037501 (1999).

\end{references}
\end{document}